\documentclass[twocolumn,floatfix,showpacs,preprintnumbers,amsmath,amssymb,superscriptaddress,prl]{revtex4-1}
\usepackage[latin9]{inputenc}
\usepackage{array}
\usepackage{float}
\usepackage{textcomp}
\usepackage{multirow}
\usepackage{amsmath}
\usepackage{amssymb}
\usepackage{graphicx}

\makeatletter

\providecommand{\tabularnewline}{\\}

\usepackage{graphics}\usepackage{subfigure}\usepackage{longtable}\usepackage{pstricks}\usepackage{dcolumn}\usepackage{bm}

\makeatother

\begin{document}

\title{Stabilization of the tetragonal structure in (Ba$_{1-x}$Sr$_{x}$)CuSi$_{2}$O$_{6}$}

\author{Pascal Puphal}

\email{puphal@physik.uni-frankfurt.de}

\affiliation{Physikalisches Institut, Goethe-Universität Frankfurt, 60438 Frankfurt
am Main, Germany}

\author{Denis Sheptyakov}

\affiliation{Laboratory for Neutron Scattering and Imaging, Paul Scherrer Institute,
5232 Villigen, Switzerland}

\author{Natalija van Well}

\affiliation{Physikalisches Institut, Goethe-Universität Frankfurt, 60438 Frankfurt
am Main, Germany}

\affiliation{Laboratory for Neutron Scattering and Imaging, Paul Scherrer Institute,
5232 Villigen, Switzerland}

\author{Lars Postulka}

\affiliation{Physikalisches Institut, Goethe-Universität Frankfurt, 60438 Frankfurt
am Main, Germany}

\author{Ivo Heinmaa}

\affiliation{National Institute of Chemical Physics and Biophysics, 12618 Tallinn,
Estonia}

\author{Franz Ritter}

\affiliation{Physikalisches Institut, Goethe-Universität Frankfurt, 60438 Frankfurt
am Main, Germany}

\author{Wolf Assmus}

\affiliation{Physikalisches Institut, Goethe-Universität Frankfurt, 60438 Frankfurt
am Main, Germany}

\author{Bernd Wolf}

\affiliation{Physikalisches Institut, Goethe-Universität Frankfurt, 60438 Frankfurt
am Main, Germany}

\author{Michael Lang}

\affiliation{Physikalisches Institut, Goethe-Universität Frankfurt, 60438 Frankfurt
am Main, Germany}

\author{Harald O. Jeschke}

\affiliation{Institut für Theoretische Physik, Goethe-Universität Frankfurt, 60438
Frankfurt am Main, Germany}

\author{Roser Valent\'{i}}

\affiliation{Institut für Theoretische Physik, Goethe-Universität Frankfurt, 60438
Frankfurt am Main, Germany}

\author{Raivo Stern}

\affiliation{National Institute of Chemical Physics and Biophysics, 12618 Tallinn,
Estonia}

\author{Christian Rüegg}

\affiliation{Laboratory for Neutron Scattering and Imaging, Paul Scherrer Institute,
5232 Villigen, Switzerland}

\affiliation{Department of Quantum Matter Physics, University of Geneva, 1205
Geneva, Switzerland }

\author{Cornelius Krellner}

\affiliation{Physikalisches Institut, Goethe-Universität Frankfurt, 60438 Frankfurt
am Main, Germany}
\begin{abstract}
{We present a structural analysis of the substituted system (Ba$_{1-x}$Sr$_{x}$)CuSi$_{2}$O$_{6}$,
which reveals a stable tetragonal crystal structure down to 1.5\,K.
We explore the structural details with low-temperature neutron and
synchrotron powder diffraction, room-temperature and cryogenic high-resolution
NMR, as well as magnetic- and specific-heat measurements and verify
that a structural phase transition into the orthorhombic structure
which occurs in the parent compound BaCuSi$_{2}$O$_{6}$, is absent
for the $x=0.1$ sample. Furthermore, synchrotron powder-diffraction
patterns show a reduction of the unit cell for $x=0.1$ and magnetic
measurements prove that the Cu-dimers are preserved, yet with a slightly
reduced intradimer coupling $J_{intra}$. Pulse-field magnetization
measurements reveal the emergence of a field-induced ordered state,
tantamount to Bose-Einstein-condensation (BEC) of triplons, within
the tetragonal crystal structure of $I\,4_{1}/acd$. This material
offers the opportunity to study the critical properties of triplon
condensation in a simple crystal structure. } 
\end{abstract}
\maketitle

\section{Introduction}

\begin{figure}[htb]
\includegraphics[width=1\linewidth]{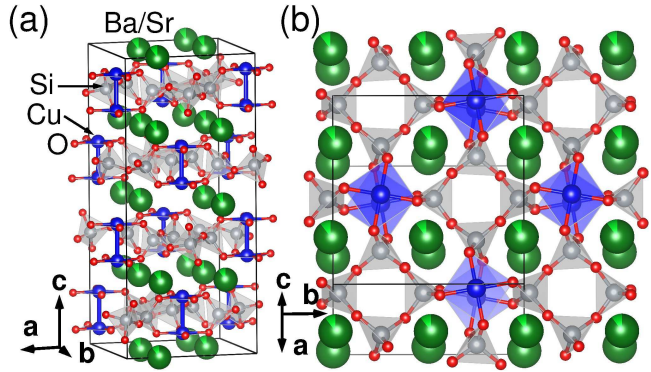}

\caption{(Ba$_{1-x}$Sr$_{x}$)CuSi$_{2}$O$_{6}$ structure for $x=0.1$.
(a) General view showing the arrangement of the Cu-dimers (blue) and
the SiO$_{4}$ tetrahedrons (grey) (b) View close to the c-axis depicting
the square lattice arrangement of the dimer layers, with highlighted
CuO$_{4}$ plaquets (blue).}

\label{fig:structure} 
\end{figure}

Magnetic insulators with Cu$^{2+}$ dimers are suitable materials
to study quantum many-body effects under variable conditions. The
occurrence of magnetic field-induced ordered states, which can be
described as Bose-Einstein condensation (BEC) of triplons in this
type of compounds provide a platform to study this ordered state in
great detail, e.g., by investigating scaling laws of thermodynamic
quantities \cite{Zapf2014}. The main idea behind this is that dimers
of two Cu$^{2+}$-ions, which each carry a spin 1/2, can be mapped
onto bosons to realize a BEC \cite{GRT2008}. A prominent material
where the appearance of a field-induced BEC of triplons was reported,
is BaCuSi$_{2}$O$_{6}$ \cite{Jaime2004}, owing its particular structure
to layers of closed rings of SiO$_{4}$ tetrahedra bridged by vertically
arranged Cu$^{2+}$ dimers which form a square lattice (see Fig.~\ref{fig:structure}).
It was proposed that frustrated inter-dimer couplings between the
dimer layers lead to a dimensional crossover at the quantum phase
transition from a paramagnetic to a field-induced magnetically ordered
state \cite{Sebastian2006}. However, it is known since 2006, that
BaCuSi$_{2}$O$_{6}$ undergoes a first-order structural phase transition
at $T\sim100K$ from a high-temperature tetragonal to a low-temperature
orthorhombic symmetry, followed by a weak incommensurability of the
crystal structure \cite{Samulon2006,Sheptyakov2012,Kraemer2007} which
thus leads to two different kinds of dimers in adjacent layers. The
impact of these two types of dimers on the peculiar properties of
the reported BEC of triplons and the role of the frustration in this
material are still under debate \cite{Sheptyakov2012,Laflorencie2009,Kr=0000E4mer2012}.

Mazurenko \textit{et al.}~\cite{Mazurenko2014} showed by performing
density functional theory (DFT) calculations based on low-temperature
structural data of the orthorhombic crystal structure, that the frustration
between dimer layers is released due to the presence of a significant
antiferromagnetic interaction between the upper site of one dimer
and the bottom site of the neighbor dimer. Such a finding, backed
by elastic neutron scattering data, questioned existing theories based
on the presence of interlayer frustration. Recently, low-temperature
high-resolution NMR (nuclear magnetic resonance) measurements~\cite{Stern2014}
detected broadened $^{29}$Si lines with complex line shape in the
orthorhombic phase suggesting a more complex structure than originally
thought, which complicates the understanding of the observed field-induced
BEC of triplons at low temperatures.

In view of the existing controversy, we follow in this work a different
strategy to avoid the influence of the structural phase transition.
We present results of a successful partial substitution of Ba by Sr
which reveals to a stable tetragonal phase of BaCuSi$_{2}$O$_{6}$
down to the lowest temperature of our experiment of 1.5\,K. With
only one type of dimers in the structure and the absence of structural
modulations down to lowest temperatures, such systems allow for the
investigation of the critical properties of field-induced ordered
states without having to deal with complications from the crystal
structure. We further present a detailed characterization of Ba$_{1-x}$Sr$_{x}$CuSi$_{2}$O$_{6}$
based on synchrotron and neutron diffraction measurements, NMR, thermodynamic
measurements and DFT calculations and show that magnetization and
susceptibility measurements for Ba$_{1-x}$Sr$_{x}$CuSi$_{2}$O$_{6}$
at $x=0.1$ display a field-induced ordered state around 22 T.

\section{Experimental details}

Polycrystalline (Ba$_{1-x}$Sr$_{x}$)CuSi$_{2}$O$_{6}$ powder samples
were prepared by sintering BaCO$_{3}$, SrCO$_{3}$, CuO and SiO$_{2}$
where the initial weight percentage of BaCO$_{3}$ was substituted
by 5, 10, 20 and 30\% SrCO$_{3}$. The powder was ground and sintered
in an aluminum oxide crucible in air at 1029\textdegree{}C (for 5\%),
1028\textdegree{}C (for 10\%), 1025\textdegree{}C and 1020\textdegree{}C
for 2 months with several steps of grinding in between. Even after
these long sintering times there is still a small amount ($<5$\%)
of impurities, which are the competing pre-phase BaCu$_{2}$Si$_{2}$O$_{7}$
and end-phase BaCuSi$_{4}$O$_{10}$, in the silicate formation \cite{Berke2007}.
We could not manage to synthesize compounds with Sr contents higher
than 30\%, but it also has to be noted that the pure SrCuSi$_{2}$O$_{6}$
phase has not been reported so far. Also a polycrystalline BaCu(Si$_{1-y}$Ge$_{y}$)$_{2}$O$_{6}$
powder sample with 10\% of SiO$_{2}$ substituted by GeO$_{2}$ was
obtained by sintering at 1027\textdegree{}C.

Single crystals were grown with pre-sintered (Ba$_{1-x}$Sr$_{x}$)CuSi$_{2}$O$_{6}$
powder spread in a boat-shaped platinum crucible. This crucible was
placed in a tube furnace with an oxygen atmosphere of 1 bar where
a viscous melt is reached at a temperature of 1150\textdegree{}C,
followed by crystallization using a cooling rate of 12 K/h. Here the
oxygen is used to suppress the decay of copper oxide $2CuO\rightarrow Cu_{2}O+\frac{1}{2}O_{2}$
as described previously \cite{Jaime2004}. The details of the crystal
growth and the influence of oxygen atmosphere on the crystal structure
will be reported elsewhere \cite{Van Well2016}.

The pure BaCuSi$_{2}$O$_{6}$ crystals were grown in a platinum crucible
with KBO$_{2}$ flux and a molar ratio of 1 : 2 (flux : powder) at
950 \textdegree{}C, where similar to \cite{SebastianPRB2006}, the
crystallization starts as a consequence of an oversaturation caused
by evaporation and some crawling due to a wetting of the crucible.

The powder diffraction experiments were carried out with two different
diffraction techniques: the high-resolution powder neutron diffractometer
HRPT \cite{Fischer2000} at the spallation neutron source SINQ and
the Powder Diffraction station of the Materials Sciences Beamline
(MS-PD) \cite{PSI} at the Swiss Light Source, both at the Paul Scherrer
Institute in Villigen. For the HRPT experiments, an amount of $\sim1$
g of Ba$_{0.9}$Sr$_{0.1}$CuSi$_{2}$O$_{6}$ was enclosed into a
vanadium can with an inner diameter of 6\,mm and the measurement
was carried out at room temperature, as well as at 1.5\,K in a $^{4}$He
bath cryostat. 

The synchrotron X-ray diffraction data were collected with the SLS-MS
Powder Diffractometer on a powder sample enclosed in a capillary with
a diameter of 0.3 mm, which was placed in a Janis flow-type cryostat
(4 - 300\,K). The Microstrip Mythen-II detector was used, which allowed
for high counting rates while maintaining the high resolution which
was essentially sample-conditioned. The typical counts of $\sim2\cdot10^{5}$
in the strongest peaks were achieved within $\sim1$ minute.

The high-resolution powder NMR-spectra were recorded with a Bruker
AVANCE-II spectrometer attached to a 8.45\,T magnet using home built
MAS-NMR probes for 1.8\,mm rotors at the National Institute of Chemical
Physics and Biophysics in Tallinn. At room temperature spectra were
recorded at 35\,kHz sample spinning speed and at low temperatures
they were recorded at about 30\,kHz. At fast magic angle spinning
(MAS) the broad NMR line of a powder sample transforms into the single
peak at the isotropic value of the magnetic shift interaction \cite{Andrew 1959}.
If the spinning speed is less than the magnetic shift anisotropy in
frequency units, then the main peak is accompanied by a number of
spinning sidebands at multiples of the spinning speed value from the
main peak. Although, the pattern of many spinning sidebands seems
complicated, it tells us unambiguously how many inequivalent nuclear
sites exist in the structure. The main purpose for using this technique
here is to show, that there is only one silicon site in doped BaCuSi$_{2}$O$_{6}$
at room temperature and at low temperature as well, whereas the $^{29}$Si
MAS NMR spectrum of the parent compound shows the appearance of additional
$^{29}$Si resonance lines below $T<100$~K \cite{Stern2014}.

Specific heat data were taken by using the standard option of a Physical
Property Measurement System (PPMS) of Quantum Design with a high heating
pulse technique discussed in chapter V.

The magnetic properties of several single crystals and a powder sample
were determined in the temperature range 2 K $\leq T\leq300$ K and
in magnetic fields $B\leq$ 5 T using a Quantum Design SQUID magnetometer.
All data have been corrected for the temperature-independent diamagnetic
core contribution of the constituents according to \cite{Kahn1993}
and the magnetic contribution of the sample holder.

High-resolution magnetization measurements were performed in a capacitor-driven
pulse-field setup with which experiments can be performed up to 58\,T
with a pulse duration of 21\,ms. The setup was equipped with a $^{4}$He-bath
cryostat. The sample was placed in a 1266 stycast can with a diameter
of 3\,mm.

\section{Structural characterization at room temperature}

\begin{figure}[H]
\includegraphics[width=1\columnwidth]{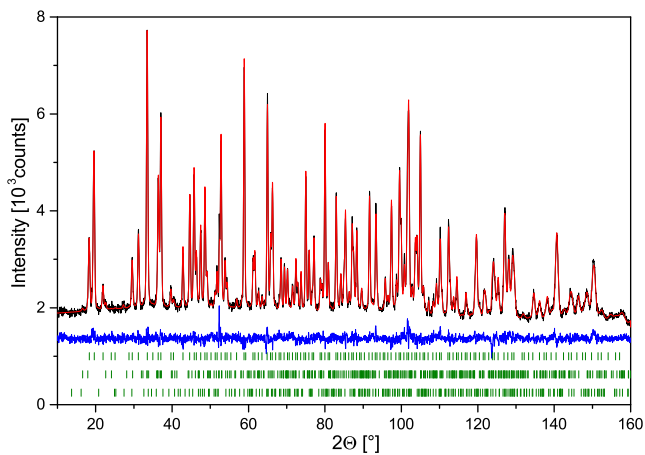} \caption{\label{neutron}Rietveld refinement of the crystal structure parameters
of (Ba$_{1-x}$Sr$_{x}$)CuSi$_{2}$O$_{6}$ compound with a $x=0.1$,
based on neutron powder diffraction data at 300~K. The observed intensity
(black), calculated profile (red), and difference curve (blue) are
shown. The rows of ticks at the bottom correspond to the calculated
diffraction peak positions of the phases (from top to bottom): BaCuSi$_{2}$O$_{6}$,
BaCu$_{2}$Si$_{2}$O$_{7}$ (2.7\% wt.) and BaCuSi$_{4}$O$_{10}$
(1.5\% wt.).}
\end{figure}

We performed a refinement of the crystal structure parameters at room
temperature by a combined analysis of neutron and synchrotron X-ray
powder diffraction data. The neutron diffraction HRPT results on 1
g of Ba$_{0.9}$Sr$_{0.1}$CuSi$_{2}$O$_{6}$ powder are shown for
the room temperature measurement in Fig.~\ref{neutron} (black curve).
The underlying refinement of the tetragonal $I\,4_{1}/acd$ structure
(red curve) fits well the measured data. Furthermore, we find a good
agreement between the neutron data and the synchrotron data as shown
in Table \ref{results} ($x=0.1$).

\begingroup 
\begin{table*}
\caption{\label{results} Summary of the refinement and EDX results of several
samples of the series (Ba$_{1-x}$Sr$_{x}$)CuSi$_{2}$O$_{6}$ and
a sample of BaCu(Si$_{1-y}$Ge$_{y}$)$_{2}$O$_{6}$. The abbreviations
are as following: neutron diffraction (N), synchrotron diffraction
(S), polycrystalline powder or crushed crystal (P), and single crystal
sample (SC).}

\begin{ruledtabular}%
\begin{tabular}{c|c|c|c|c|c|c|c|c|c}
{\footnotesize{{{{{Nominal value} }}}}}  & {\footnotesize{{{{{$x=0.05$} }}}}}  & \multicolumn{5}{c|}{{\footnotesize{{{{{$x=0.1$}}}}}}} & {\footnotesize{{{{{$x=0.2$} }}}}}  & {\footnotesize{{{{{$x=0.3$} }}}}}  & {\footnotesize{{{{{$y=0.1$}}}}}}\tabularnewline
\hline 
{\footnotesize{{{{$\begin{array}{c}
\mbox{Powder/}\\
\mbox{Single crystal}
\end{array}$}}}}}  & {\footnotesize{{{{{P} }}}}}  & \multicolumn{2}{c|}{{\footnotesize{{{{{SC}}}}}}} & \multicolumn{3}{c|}{{\footnotesize{{{{{P} }}}}}} & {\footnotesize{{{{{P} }}}}}  & {\footnotesize{{{{{P} }}}}}  & {\footnotesize{{{{{P}}}}}}\tabularnewline
{\footnotesize{{{{{$x_{\text{EDX}}$ value} }}}}}  & {\footnotesize{{{{{0.08(1)} }}}}}  & \multicolumn{2}{c|}{{\footnotesize{{{{{0.08(2)}}}}}}} & \multicolumn{3}{c|}{{\footnotesize{{{{0.13(1)}}}}}} & {\footnotesize{{{{{0.19(2)} }}}}}  & {\footnotesize{{{{{0.33(3)} }}}}}  & {\footnotesize{{{{{0.08(2)}}}}}}\tabularnewline
\hline 
{\footnotesize{{{{$\begin{array}{c}
\mbox{Neutron/}\\
\mbox{Synchrotron}
\end{array}$}}}}}  & {\footnotesize{{{{{S} }}}}} & \multicolumn{2}{c|}{{\footnotesize{{{{{S}}}} \cite{Appendix}}}} & \multicolumn{2}{c|}{{\footnotesize{{{{{N} }}} \cite{Appendix}}}} & {\footnotesize{{{{{S} }}}}}  & {\footnotesize{{{{{S} }}}}}  & {\footnotesize{{{{{S} }}}}}  & {\footnotesize{{{{{S}}}}}}\tabularnewline
{\footnotesize{{{{{$x_{\text{refined}}$ value} }}}}}  & {\footnotesize{{{{{0.03} }}}}}  & \multicolumn{2}{c|}{{\footnotesize{{{{{0.05}}}}}}} & \multicolumn{2}{c|}{{\footnotesize{{{{{0.07} }}}}}} & {\footnotesize{{{{{0.09} }}}}}  & {\footnotesize{{{{{0.16} }}}}}  & {\footnotesize{{{{{0.26} }}}}}  & {\footnotesize{{{{{0.08}}}}}}\tabularnewline
\hline 
{\footnotesize{{{{Temperature }}}}}  & {\footnotesize{{{{295\,K}}}}}  & {\footnotesize{{{{295\,K}}}}}  & {\footnotesize{{{{{4\,K} }}}}}  & {\footnotesize{{{{{300\,K} }}}}}  & {\footnotesize{{{{1.5\,K}}}}}  & {\footnotesize{{{{295\,K}}}}}  & {\footnotesize{{{{295\,K}}}}}  & {\footnotesize{{{{295\,K}}}}}  & {\footnotesize{{{{295\,K}}}}}\tabularnewline
{\footnotesize{{{{{$a$ {[}{\AA{}}{]}} }}}}}  & {\footnotesize{{{{{9.97331(2)} }}}}}  & {\footnotesize{{{{{9.97223(2)} }}}}}  & {\footnotesize{{{{{9.95830(5)} }}}}}  & {\footnotesize{{{{{9.9627(3)} }}}}}  & {\footnotesize{{{{9.9508(2)}}}}}  & {\footnotesize{{{{{9.95888(2)} }}}}}  & {\footnotesize{{{{{9.94442(1)} }}}}}  & {\footnotesize{{{{{9.93810(3)} }}}}}  & {\footnotesize{{{{{9.9935(1)}}}}}}\tabularnewline
{\footnotesize{{{{{$c$ {[}{\AA{}}{]}} }}}}}  & {\footnotesize{{{{{22.30826(6)} }}}}}  & {\footnotesize{{{{{22.31379(4)} }}}}}  & {\footnotesize{{{{{22.3246(1)} }}}}}  & {\footnotesize{{{{{22.2774(7)} }}}}}  & {\footnotesize{{{{22.2815(5)}}}}}  & {\footnotesize{{{{{22.27168(6)} }}}}}  & {\footnotesize{{{{{22.23223(5)} }}}}}  & {\footnotesize{{{{{22.2129(1)} }}}}}  & {\footnotesize{{{{{22.4325(2)}}}}}}\tabularnewline
{\footnotesize{{{{{$V$ {[}{\AA{}}$^{3}${]}} }}}}}  & {\footnotesize{{{{{2218.93} }}}}}  & {\footnotesize{{{{{2219.08} }}}}}  & {\footnotesize{{{{{2213.88} }}}}}  & {\footnotesize{{{{{2210.58} }}}}}  & {\footnotesize{{{{2205.31}}}}}  & {\footnotesize{{{{{2208.89} }}}}}  & {\footnotesize{{{{{2198.58} }}}}}  & {\footnotesize{{{{{2193.88} }}}}}  & {\footnotesize{{{{{2240.33}}}}}}\tabularnewline
\end{tabular}\end{ruledtabular} 
\end{table*}

\endgroup

\begin{figure}[h]
\includegraphics[width=1\columnwidth]{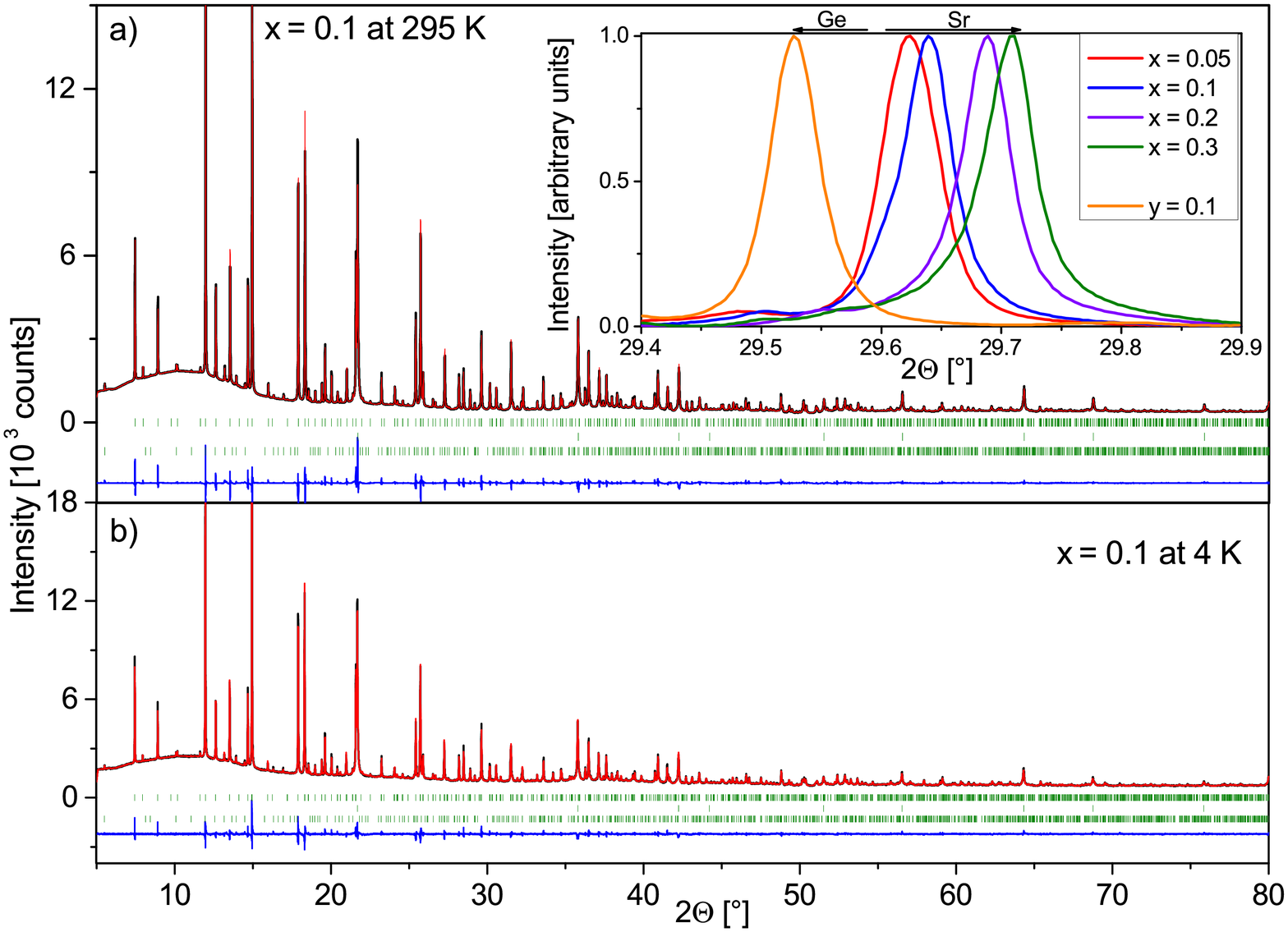} 

\caption{\label{refined}Rietveld refinement of the crystal structure parameters
of (Ba$_{1-x}$Sr$_{x}$)CuSi$_{2}$O$_{6}$ crushed single crystal
with a $x=0.1$, based on synchrotron X-ray powder diffraction data
at a) 295~K and b) 4~K. The rows of ticks in the middle correspond
to the calculated diffraction peak positions of the phases (from top
to bottom): BaCuSi$_{2}$O$_{6}$, diamond powder (added to reduce
the absorption) and BaCuSi$_{4}$O$_{10}$ (3.4 \% wt.). The inset
shows synchrotron X-ray powder diffraction data of the $(624)$ peak
measured at 295\,K of polycrystalline samples with various substitution
levels.}
\end{figure}

In addition to the powder data, a crushed single crystal of Ba$_{0.9}$Sr$_{0.1}$CuSi$_{2}$O$_{6}$
was measured with synchrotron X-ray diffraction at room temperature
and the measured data together with the refinement is shown in Fig.~\ref{refined}.
We observe that the tetragonal structure is preserved at doping levels
up to 30\% at room temperature. The unit cell dimensions decrease
with increasing strontium amount. This can be seen in a shift of the
diffraction peaks to higher angles (see the inset in Fig.~\ref{refined})
and in the evolution of the unit cell parameters in Table \ref{results}.
This continuous evolution evidences that the Sr is built in, replacing
the larger Ba atoms. In addition, the synchrotron data in the inset
of Fig. \ref{refined} indirectly indicate a homogeneous distribution
of strontium in the material, since the FWHM (full-width at half minimum)
values show no significant increase with increasing strontium content.
For example, the FWHM\textquoteright{}s of the (624) peak are 0.046\textdegree{},
0.050\textdegree{}, 0.049\textdegree{} and 0.051\textdegree{} for
compounds with x = 0.05, 0.1, 0.2 and 0.3 respectively. Comparing
neutron data from a BaCuSi$_{2}$O$_{6}$ powder to the $x=0.1$ powder
we see that introducing Sr into the structure causes a slight peak
broadening, which is qualitatively indicative to the presence of microstrains
in the substituted materials. A further effect is that a decreased
thermal expansion is observed, which influences the Cu-Cu distances
and the different exchange couplings J at low temperatures in Table
\ref{tab:couplings}.

The results of all refinements and of the energy dispersive X-ray
spectroscopy (EDX) measurements are presented in Table 1 and it is
apparent, that the actual Sr content slightly varies from sample to
sample. As a general trend, the amount of strontium, $x$, in the
powder is lower than the nominal one and this value further decreases
in the single crystals. The occupation refinement value $x$ is compared
to the data obtained from the EDX analysis in a Zeiss DSM 940A scanning
electron microscope (SEM) on both powder and single crystals. The
Sr content from the EDX measurement on the powder samples seems to
be slightly overestimated compared to the value from the refinement.
The reason for this is possibly related to Sr-enriched foreign phases,
which increase the amount of Sr in the EDX result, while in the refined
results they are refined seperately. Also SrO impurities cannot be
detected in diffraction experiments, since it decays and becomes amorphous.
For clarity, we will use in the following the nominal values to describe
the samples.

\begin{figure}[h]
\includegraphics[width=0.9\columnwidth]{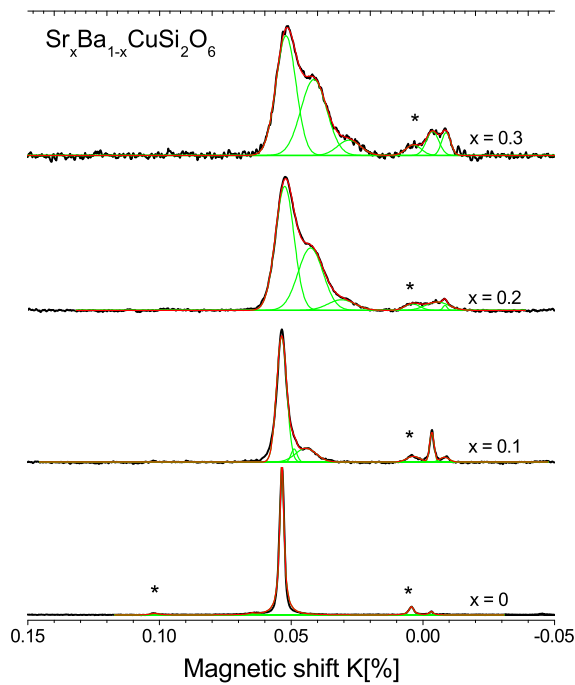} \caption{\label{NMR} $^{29}$Si MAS-NMR spectra at $T=300$\,K with a resonance
frequency of 71.5 MHz of $^{29}$Si on BaCuSi$_{2}$O$_{6}$ (bottom
spectrum) and Sr substituted BaCuSi$_{2}$O$_{6}$ as noted in the
figure. The asterisks denote spinning side bands. Decomposition of
the spectra can be seen by analyzing the green fitting curves. Homogeneous
substitution of Ba by Sr is clearly reflected in the $^{29}$Si MAS-NMR
spectra.}
\end{figure}

To have a further insight into the distribution of Sr in the samples,
room-temperature, high-resolution $^{29}$Si-NMR measurements were
done on the Sr-powder substitution series. The results of the chemical
shift $K$ are depicted in Fig.~\ref{NMR} and a broadening of the
main peak around $K\sim0.05$\% is apparent. This line broadening
can be explained, assuming a random distribution of Sr on the Ba sites.
In BaCuSi$_{2}$O$_{6}$, each silicon atom, which is the probed NMR
nuclei, has two Ba nearest neighbors. For the $x=0.1$ compound the
probability to have a silicon atom with two Ba neighbors is 0.81,
with one Ba and one Sr 0.18, and with two Sr neighbors only 0.01,
if Sr is homogeneously distributed over the Ba sites. Therefore, one
would expect a peak splitting of the main $^{29}$Si-NMR line into
a triplet with an intensity ratio of (0.81 : 0.18 : 0.01). Looking
at the spectra in Fig.~\ref{NMR}, such a peak splitting is indeed
observed. At 10\% Sr concentration the spectrum shows the main line
at 0.0536\,\% and a shoulder at 0.0444\,\%. The intensity ratio
of these lines is (0.83 : 0.17), which is in nice agreement, with
the expected splitting, although the peak with two Sr sites is below
the detection limit.

The main lines in the spectrum of 20\% Sr are positioned at 0.0526\,\%,
0.0427\,\% and 0.0310\,\% with an intensity ratio of (0.55 : 0.38
: 0.07), in rough correspondence with the expected site distribution,
which would lead to a side distribution of (0.64 : 0.32 : 0.04). The
main lines in the spectrum of 30\% Sr substituted compound are at
0.0520\,\%, 0.0414\,\% and 0.0280\,\%. The intensity distribution
of the lines (0.51 : 0.42 : 0.07) is again in good agreement with
a random distribution of Sr sites (0.49 : 0.42 : 0.09).

\section{Investigation of the absence of a structural phase transition}

We discuss now the low-temperature diffraction data obtained both
in measurements on powder as well as single crystal samples with the
nominal Sr concentration $x=0.1$. The absence of a structural phase
transition down to the lowest measured temperatures could be verified
in neutron and synchrotron diffraction. In Fig.~\ref{refined} b)
the Rietveld refinement of the same crushed single crystal as in Fig.~\ref{refined}
a) is presented for measurements at 4 K. This data set can be refined,
using the same tetragonal crystal structure with space group $I\,4_{1}/acd$,
which is observed also at room temperature. The corresponding results
of the refinement are shown in Table \ref{results}.

\begin{figure}[h]
\includegraphics[width=1\columnwidth]{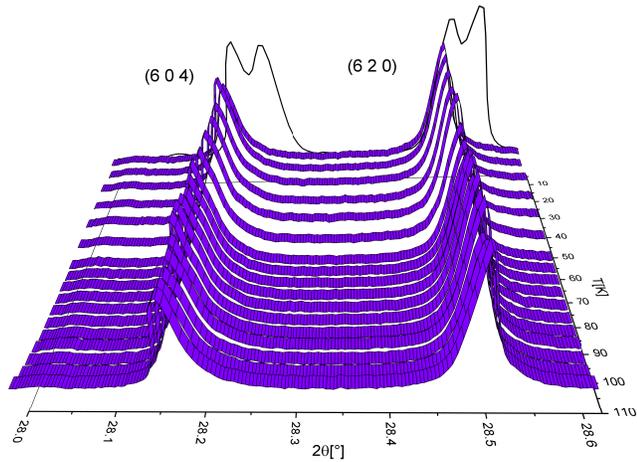} \caption{\label{PT} Synchrotron X-ray powder diffraction data of an $x=0.1$
crushed single crystal from 35 to 110\,K. In the back a measurement
of the $x=0$ powder at 30~K is shown (black line), demonstrating
how the peak splitting due to the transition would look like (data
taken from of Ref. \cite{Sheptyakov2012}). The suppression of the
structural phase transition is apparent, since for Ba$_{0.9}$Sr$_{0.1}$CuSi$_{2}$O$_{6}$
the (604) and (620) reflexes do not split.}
\end{figure}

Further evidence for the absence of a structural phase transition
is presented in Fig.~\ref{PT}, where the temperature-dependent synchrotron
data across the expected transition temperature are shown. We choose
the (604) and (620) reflexes in a temperature window from 10 to 110\,K
to make a comparison with Ref.~\cite{Sheptyakov2012} possible, where
corresponding data were shown for $x=0$. There, a well resolved peak
splitting is observed as shown for comparison in the back of Fig.
\ref{PT}. In the whole temperature region for the $x=0.1$ sample,
no peak splittings or shifts beyond the expected thermal expansion
could be observed. Similar results were obtained with neutron diffraction
at 1.5\,K, where on a polycrystalline sample, with slightly higher
Sr concentration, also no structural phase transition could be detected.
The same suppression of the phase transition was observed for a 10
\% Ge doped powder sample ($y=0.1$ in Table \ref{results}) in synchrotron
measurements in the range of 4~K $<T<$ 295\,K.

\begin{figure}[h]
\includegraphics[width=1\columnwidth]{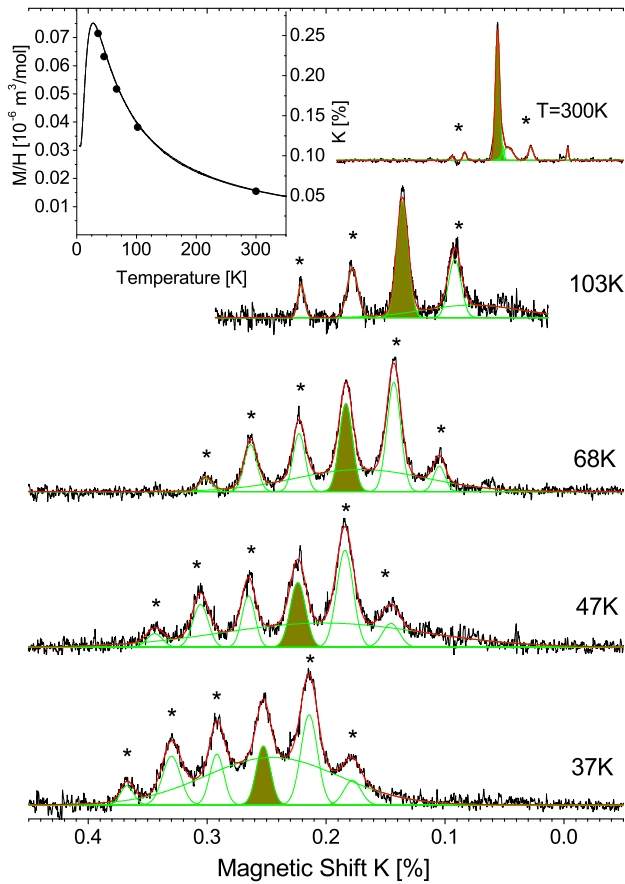} \caption{\label{cryoMAS} The temperature dependence of $^{29}$Si MAS-NMR
spectrum of Ba$_{0.9}$Sr$_{0.1}$CuSi$_{2}$O$_{6}$. At magic-angle
spinning the NMR spectra consist of the main line at isotropic magnetic
shift and of a number of spinning sidebands at multiples of the sample-spinning
frequency from the main line. For clarity the main lines in the spectra
are colored and the spinning sidebands are marked with asterisks.
The insert shows the proportionality of the isotropic magnetic shift
to the molar susceptibility measured in a PPMS on the same sample
at 8.45~T. Here the susceptibility values are given by the full line
and circles correspond to the isotropic shift values.}
\end{figure}

Next to the scattering experiments (sensitive to long-range structures),
we also performed cryogenic high-resolution $^{29}$Si NMR which show
the absence of the transition in short-range correlations. The spectra
show that, unlike the case of the parent compound BaCuSi$_{2}$O$_{6}$,
where two different $^{29}$Si resonance bands were found below the
phase transition at T $\sim$96~K \cite{Stern2014}, there is clearly
only one resonance line in the studied temperature range 37~K $\leq$
T $\leq$ 300~K (see figure \ref{cryoMAS}). With lower temperatures
there is a natural line broadening due to high magnetic suscebtibility
of the powder particles which is not averaged by MAS, making the structure
(shoulders) caused by Sr less and less detectable. The isotropic value
of the $^{29}$Si magnetic hyperfine shift in the $x=0.1$ sample
at room temperature for silicon with two Ba neighbors $K=0.0536$~\%
is equal to the value in pure BaCuSi$_{2}$O$_{6}$ ($K=0.0535$~\%).
In paramagnetic compounds the isotropic magnetic shift K is proportional
to the magnetic suscebtibility $\chi_{M}$ as: $K=\frac{H_{hf}}{N_{A}g\mu_{B}}\chi_{M}$
where $H_{hf}$ is the hyperfine field at the nucleus, $N_{A}$ is
the Avogadro's number, g and $\mu_{B}$ are the g-factor and the
Bohr magneton, respectively. Equal magnetic hyperfine shifts result
from equal hyperfine fields at silicon in {\small{{Ba$_{0.9}$Sr$_{0.1}$CuSi$_{2}$O$_{6}$}}}
and in the parent compound {\small{{BaCuSi$_{2}$O$_{6}$}.}}{\small \par}

\begin{figure}[h]
\includegraphics[width=1\columnwidth]{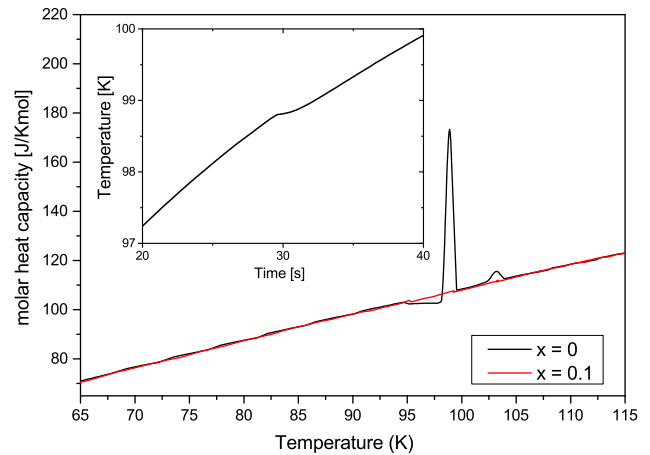}

\caption{\label{spec heat} Specific heat data of a BaCuSi$_{2}$O$_{6}$ single
crystal grown in KBO$_{2}$ flux as well as a Ba$_{0.9}$Sr$_{0.1}$CuSi$_{2}$O$_{6}$
single crystal grown with oxygen partial pressure. The inset shows
a single heat pulse in the vicinity of the structural phase transition
of the BaCuSi$_{2}$O$_{6}$ single crystal.}
\end{figure}

A complementary measurement technique to detect first-order structural
phase transitions is the heat capacity measured around the suspected
phase transition. The advantages of this method are that the measurement
is fast, the single crystals are kept intact and can be small. We
measured specific heat data from 10 to 130~K with heating pulses
of up to 10~K of single crystals with and without strontium substitution.
As a consequence of the first-order nature of the structural transition
in pure BaCuSi$_{2}$O$_{6}$, latent heat is expected, which easily
can be detected as a small plateau during a continuous heating cycle
(inset of Fig.~\ref{spec heat}) as described in Ref. \cite{Lashley2003}.
This results in a diverging specific heat at the transition temperature
as evident from the large peak in the main part of Fig.~\ref{spec heat}
(black curve). As this is a first-order transition, we observe a small
satellite peak at higher temperatures, due to hysteresis effects upon
heating and cooling. This anomaly is found to be absent for Sr-substituted
samples, which is exemplarily shown for one $x=0.1$ single crystal
in Fig.~\ref{spec heat} (red curve).

\section{Magnetic characterization}

After having established that in Sr-substituted BaCuSi$_{2}$O$_{6}$
the structural transition into the orthorhombic structure is suppressed,
the question arises, how these structural differences influence the
magnetic properties at low temperatures and high magnetic fields.
Here, we present magnetic susceptibility measurements on a Ba$_{0.9}$Sr$_{0.1}$CuSi$_{2}$O$_{6}$
single crystal down to 2\,K together with high-field magnetization
measurements up to 50\,T at 1.5\,K. For comparision, we also studied
a single crystal of the undoped parent compound. 

Using a SQUID magnetometer we determine the molar magnetic susceptibility
of a BaCuSi$_{2}$O$_{6}$ single crystal of 8.79 mg. Futhermore we
measured two $x=0.1$ single crystals with masses of 4.07\,mg (\#1)
and 5.29\,mg (\#2) and a Ba$_{0.9}$Sr$_{0.1}$CuSi$_{2}$O$_{6}$
powder sample of 110\,mg in the temperature range 2\,K $\leq T\leq$
300\,K in a field of 1\,T. A powder sample of this size ensures
an optimal filling factor of the pick-up coil in the pulse-field experiments.
In addition, we determine the magnetization of the powder up to 5~T
at 2~K. The single crystals and the powder sample are of high quality
as reflected by low paramagnetic (spin-1/2) impurity levels of 0.8
\% for \#1, 1.5 \% for \#2 and 5.5 \% for the powder. For the undoped
single crystal this spin-1/2 impurity level amounts to about 1.0 \%.
These impurities might arise from paramagnetic BaCuSi$_{4}$O$_{10}$
\cite{Masunaga2015}, observed in our x-ray and neutron diffraction
data.

\begin{figure}[h]
\includegraphics[width=1\columnwidth]{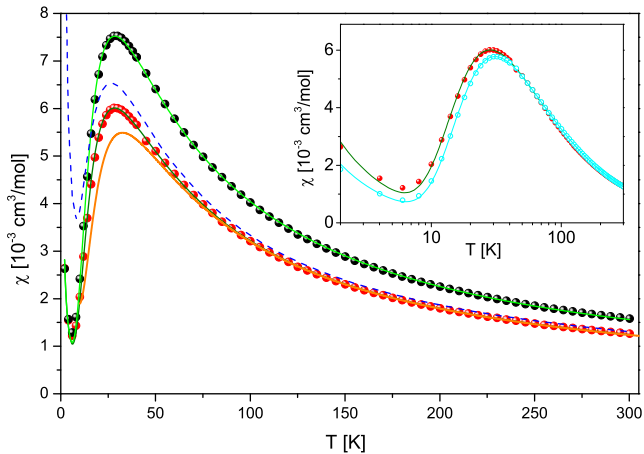} \caption{\label{SQUID} Molar susceptibility of single crystal \#2 for $B_{\bot c}$
(full red circles) and $B_{\parallel c}$ (full black circles) together
with $\chi_{m}$ of the powder sample (blue broken line) measured
as a function of temperature. Due to the random orientation of the
micro crystals in the powder, its susceptibility lies between the
data for $B_{\bot c}$ and $B_{\parallel c}$. The data of \#2 for
the different orientations are fitted with a random-phase approximation
(RPA) according to \cite{Luethi2004} together with a Curie contribution
resulting from isolated $S=\frac{1}{2}$ impurities (fits are the
full green lines). The full orange line corresponds to the expected
$\chi_{m}\left(T\right)$ using high-temperature series expansion
(HTSE) \cite{Lohmann2014} with the magnetic coupling constants taken
from the DFT calculation, see chapter VI. Inset: $\chi\left(T\right)$of
\#2 (full red circles) together with the data of an undoped single
crystal (open cyan circles) for $B_{\bot c}$. The solid lines are
fits to the data using a RPA expression given in Ref. \cite{Luethi2004}.}
\end{figure}

In a first attempt to extract the relevant magnetic coupling parameter
$J_{{\rm intra}}$ (intradimer Cu-Cu exchange) (main panel of Fig.~\ref{SQUID})
and $J_{{\rm inter}}$ (average interdimer exchange) we performed
combined fits of $\chi_{m}$ for $B_{\bot c}$ and $B_{\parallel c}$.
For the $g$-factor, treated as the only independent parameter in
fitting the two data sets, we obtained $g_{\bot c}=2.07$ and $g_{\parallel c}=2.32$.
These are the typical values for Cu$^{2+}$ ions in a square-planar
environment \cite{ottaviani}. For the antiferromagnetic intradimer
coupling constant the fit yields $J_{{\rm intra}}=46.7(5)$\, K for
\#2 together with an antiferromagnetic coupling between dimers of
$J_{{\rm inter}}=10(2)$\, K. For \#1 (not shown) a fit of comparable
quality results in slightly different values of $J_{{\rm intra}}=47.7(5)$\,K
and $J_{{\rm inter}}=8(2)$\, K. We believe that these differences
in the magnetic coupling constants are due to deviations in the actual
Sr concentrations from $x=0.1$, as discussed above. These numbers
can be compared with the results on the parent compound BaCuSi$_{2}$O$_{6}$.
From fitting the data of the undoped single crystal in the same way
as for crystals \#1 and \#2 (see the inset of Fig. \ref{SQUID} for
$B_{\bot c}$) we obtain a sightly higher value for $J_{{\rm intra}}=50.4(5)$\,K
together with $J_{{\rm inter}}=8(2)$\,K and a g-factor of 2.08.
On the semi-logarithmic scale of the inset of Fig. \ref{SQUID} this
small difference is directly visible in the shift of the maxima. Note,
that these magnetic coupling constants are similar to the ones obtained
in Ref. \cite{Jaime2004} where they were determined from the slope
of the M(H) curves at 37 T for different temperatures using a Quantum
Monte Carlo algorithm.

\begin{figure}[h]
\includegraphics[width=1\columnwidth]{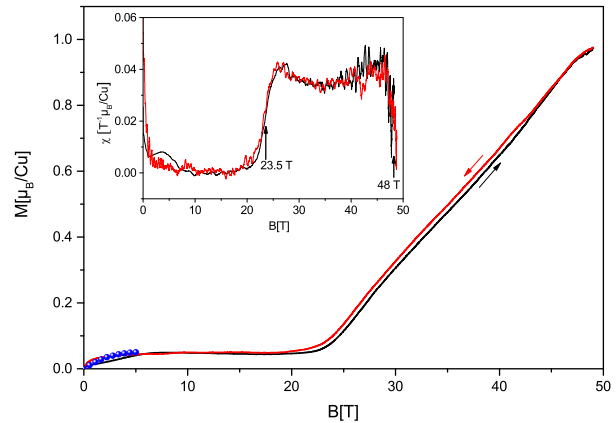} \caption{\label{pulsed} Magnetization of the Ba$_{0.9}$Sr$_{0.1}$CuSi$_{2}$O$_{6}$
powder sample up to 50\,T measured at a bath temperature of 1.5\,K.
The black solid line represents the data taken during increasing field
with a rise time of 8\,ms whereas the red solid line shows the magnetization
with decreasing field with a decay time of 17\, ms. The blue solid
points are the SQUID data taken at 2\,K. They follow a Brillouin-function
corresponding to a concentration of 5.5 \% uncoupled Cu$^{2+}$(spin-1/2)-ions.
These data are used to calibrate the pulse-field experiments and they
allow a rough estimate of the magnetocaloric effect which amounts
to $\Delta T\simeq$ +1.5\,K for the field-up curve and $\simeq$
-0.5\,K for the field-down curve with respect to the bath temperature.}
\end{figure}

In Fig.\,\ref{pulsed} we show the results of the magnetization,
$M$, as a function of magnetic field up to 50\,T at a bath temperature
of 1.5\,K. In small fields we observe a mild increase of $M$ which
levels off at intermediate fields. We assign this to the Brillouin
function of uncoupled Cu$^{2+}$(spin-1/2)-ions. By subtracting the
corresponding contribution from the raw data we find zero magnetization
up to a field of around 22\,T. With further growing field, $M(B)$
increases almost linearly with $B$, until the saturation is reached
at $B_{c2}$ around 48\,T. The small deviations between the field-up
and field-down data are due to the magnetocaloric effect (MC), i.e.,
temperature changes due to changes of the magnetic field, in combination
with the peculiar field-time characteristic of the pulse-field set
up. The largest MC is expected for fields $\leq$ 5\,T and around
the critical fields $B_{c1}$ and $B_{c2}$, with $B_{c1}$ denoting
the onset field of the field-induced order. In order to determine
$B_{c1}$, we numerically differentiate the data and obtain the magnetic
susceptibility, shown in the inset of Fig.\,\ref{pulsed}. We define
$B_{c1}$ as the inflection point of the $\chi\left(B\right)$ curve
in analogy to the criterion used in Ref.\,\cite{Jaime2004}. For
the powder sample we obtain $B_{c1}$ = 23.5\,T at a temperature
of about 2\,K. This temperature is corrected for the MC. For a stack
of single crystals (not shown here) we obtained a slightly smaller
value of 22.3\,T. Since $B{}_{c1}$ scales with $J_{intra}/g$, a
slight reduction observed in $B{}_{c1}$ for the single crystals would
be consistent with a 10\% larger $g$-factor for fields perpendicular
to the planes even though $J_{intra}$ is slightly (maximally 5\%)
enhanced. As estimated in Ref.\,\cite{Mazurenko2014} a $B_{c1}$
around 22\,T corresponds to one of the dimer layers (layer A) in
the orthorhombic low-temperature phase. This layer is structurally
similar to the dimer layers in the tetragonal I4$_{1}$/acd structure
of BaCuSi$_{2}$O$_{6}$.

\section{Electronic structure calculations}

\begin{figure}[h]
\includegraphics[width=1\columnwidth]{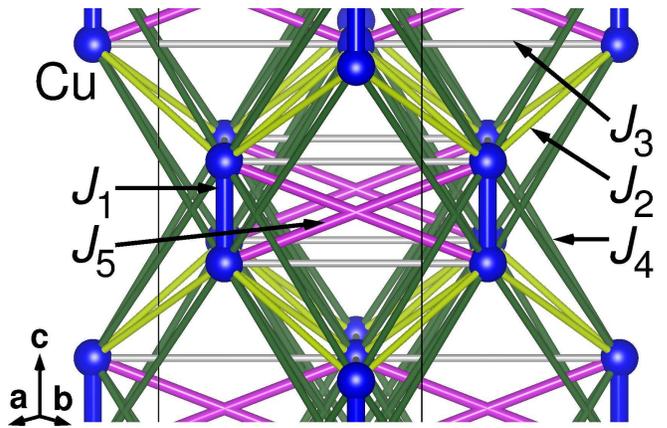} \caption{Detail of the BaCuSi$_{2}$O$_{6}$ unit cell with the first five
exchange paths between Cu$^{2+}$ ions. Other ions are omitted for
clarity.}

\label{fig:couplings} 
\end{figure}

In order to provide a more detailed analysis of the Cu-Cu interactions
in Ba$_{0.9}$Sr$_{0.1}$CuSi$_{2}$O$_{6}$ beyond the $J_{{\rm intra}}$
and $J_{{\rm inter}}$ estimates from the previous section, we perform
density functional theory calculations on the neutron and synchrotron
diffraction refinements of 10{\%} Sr doped BaCuSi$_{2}$O$_{6}$
samples (see Table \ref{results}, $x=0.1$) at room and low temperatures.
We employ the all electron full potential local orbital (FPLO) code~\cite{FPLOmethod}
using a generalized gradient approximation~\cite{PerdewBurkeErnzerhof}
exchange and correlation functional and correct for the strong correlations
on the Cu$^{2+}$ $3d$ orbitals with the GGA+U~\cite{Liechtenstein95}
functional. We lower the symmetry of (Ba$_{1-x}$Sr$_{x}$)CuSi$_{2}$O$_{6}$
from $I\,4_{1}/acd$ to $C\,2$ in order to make eight Cu sites inequivalent
and calculate the total energies of 21 distinct spin configurations
for each of the four structures. Note that the isoelectronic substitution
of 10{\%} Sr$^{2+}$ for Ba$^{2+}$ is reflected in the calculation
only by the experimentally determined lattice constants and interatomic
distances but not by actual replacement of Ba sites in a supercell
approach. The 21 energies can be fitted~\cite{Tutsch2014} against
five Heisenberg exchange couplings $J_{i}$ with very high accuracy,
leading to very small error bar from the statistics. Note that the
sub-Kelvin error bars result from the particularly well defined $S=\frac{1}{2}$
moments of Cu in BaCuSi$_{2}$O$_{6}$, leading to very precise mapping
of the 21 DFT total energies to the Hamiltonian with five exchange
couplings. The results for GGA+U interaction parameters $U=8$~eV
and $J_{H}=1$\,eV are given in Table~\ref{tab:couplings}. The
parameters $U=8$~eV and $J_{H}=1$\,eV are chosen on the upper
end of the interaction parameter range $U\in[6,8]$\,eV considered
in previous studies for Cu$^{2+}$ in square-planar oxygen environment\,\cite{HerbertsmithiteExchange,BarlowiteTheoryExperiment}.
The five exchange paths are visualized in Fig.~\ref{fig:couplings}.
$J_{{\rm intra}}$ as introduced in the previous section, corresponds
to $J_{1}$ while $J_{{\rm inter}}$ corresponds to a non-trivial
average of interdimer Cu-Cu interactions including $J_{2},J_{3},J_{4}$
and $J_{5}$. In Table~\ref{tab:couplings} we also show the calculated
exchange parameters for the $x=0$ structure at 200\,K (Ref.~\cite{Sparta2003})
and include, for comparison, the results calculated in Ref.~\cite{Mazurenko2014}
for the room-temperature tetragonal BaCuSi$_{2}$O$_{6}$. We observe
(i) a good agreement between our estimates and those of Ref.~\cite{Mazurenko2014}
for $x=0$ in the tetragonal phase and (ii) a reasonably good agreement
between our \textit{ab initio}-calculated intradimer $J_{1}$ and
$J_{{\rm intra}}$ obtained in the previous section. We use a high-temperature
series expansion~\cite{Lohmann2014} in order to check if the calculated
exchange couplings can explain the experimentally measured susceptibility.
The values we obtain for the T = 1.5~K neutron structure produce
the orange curve in Fig. \ref{SQUID}, with a maximum at 31~K in
good agreement to the experimental maximum at 32.5~K. (iii) Our calculation
of the Hamiltonian parameters for the low temperature tetragonal structure
of Ba$_{1-x}$Sr$_{x}$CuSi$_{2}$O$_{6}$ at nominal $x=0.1$ show
that the exchange interactions remain very similar to the couplings
of the $T=200$~K tetragonal structure of BaCuSi$_{2}$O$_{6}$.
Clearly Ba$_{0.9}$Sr$_{0.1}$CuSi$_{2}$O$_{6}$, as well as the
tetragonal I4$_{1}$/acd BaCuSi$_{2}$O$_{6}$ phase display strong
intradimer antiferromagnetic Cu-Cu couplings ($J_{1}$) and significant
nearest-neighbor dimer top-bottom antiferromagnetic couplings ($J_{5}$)
that release any type of possible frustration between dimer layers.

\begin{table}
\caption{\label{tab:couplings} Calculated exchange couplings for the (Ba$_{1-x}$Sr$_{x}$)CuSi$_{2}$O$_{6}$
structures with nominal $x=0.1$ given in Table~1. A GGA+U functional
with $U=8$~eV and $J_{H}=1$~eV is used. The abbreviations are
neutron diffraction (N) and synchrotron diffraction (S). In addition,
couplings for pure tetragonal BaCuSi$_{2}$O$_{6}$ (structure from
Ref.~\onlinecite{Sparta2003}) are given in the last two lines. }

\begin{ruledtabular} %
\begin{tabular}{llrrrrr}
 & T (K)  & $J_{1}$\,(K)  & $J_{2}$\,(K)  & $J_{3}$\,(K)  & $J_{4}$\,(K)  & $J_{5}$\,(K) \tabularnewline
\hline 
\multirow{2}{*}{N} & 1.5  & 51.6(1)  & -0.27(1)  & -0.41(1)  & 0.0(1)  & 7.9(1)\tabularnewline
 & 300  & 58.6(1)  & -0.27(2)  & -0.35(1)  & 0.0(1)  & 8.3(1)\tabularnewline
\multirow{2}{*}{S} & 4  & 60.2(1)  & -0.25(1)  & -0.41(1)  & 0.0(1)  & 8.0(1)\tabularnewline
 & 295  & 56.7(1)  & -0.28(4)  & -0.29(3)  & 0.0(1)  & 8.3(3)\tabularnewline
\hline 
$x=0$  & 200  & 58.7(1)  & -0.23(1)  & -0.39(1)  & 0.0(1)  & 8.4(1)\tabularnewline
$x=0$~\cite{Mazurenko2014}  & 300  & 53  & -0.3  & -0.4  & --  & 7.9 \tabularnewline
\end{tabular}\end{ruledtabular} 
\end{table}

\section{Conclusions}

We have experimentally confirmed the absence of a first-order tetragonal-to-orthorhombic
structural phase transition in (Ba$_{1-x}$Sr$_{x}$)CuSi$_{2}$O$_{6}$
by means of powder synchrotron X-ray and neutron diffraction, NMR,
thermodynamic measurements and density functional theory calculations.
We find that such a phase transition is suppressed with strontium
substitution. Furthermore, the unit-cell volume decreases with increasing
Sr content and the intradimer magnetic coupling becomes slightly reduced.
Our DFT calculations for $x=0.1$ for the tetragonal I4$_{1}$/acd
structures show the presence of strong antiferromagnetic Cu-Cu intradimer
couplings and non-negligible antiferromagnetic nearest-neighbor dimer
top-bottom antiferromagnetic couplings that avoid any kind of frustration
between the dimer layers. The fact that for the germanium substituted
sample the phase transition is also suppressed leaves us with a readily
tunable system by varying the substitution concentrations. First high-field
magnetic measurements on a powder sample with $x=0.1$ at 2\,K reveal
clear indications for a field-induced ordered state, similar to the
observations reported for the $x=0$ parent compound. In contrast
to the $x=0$ material, however, where the analysis of the critical
properties are plagued by uncertainties related to the presence of
two sorts of dimers as a consequence of the structural transition,
the $x=0.1$ material is free of this complication. Therefore, detailed
high-field measurements on this new material may help to clarify the
influence of structural subtleties on the critical behavior of the
field-induced order.
\begin{acknowledgments}
The authors gratefully acknowledge support by the Deutsche Forschungsgemeinschaft
through grant SFB/TR 49 and project WE-5803/1-1. {\small{The work
in Tallinn was supported by Estonian Research Council grants PUT${210}$
and IUT${23-7}$.}} This work is partly based on experiments performed
at the Swiss spallation neutron source SINQ, and the Swiss Light Source
synchrotron radiation source at Paul Scherrer Institute, Villigen,
Switzerland. \end{acknowledgments}

\end{document}